\documentclass[letterpaper]{article} 
\usepackage{amsmath}
\usepackage{arxiv}
\usepackage{times}  
\usepackage{helvet}  
\usepackage{courier}  
\usepackage[hyphens]{url}  
\usepackage{graphicx} 
\usepackage{natbib}  
\usepackage{caption} 
\usepackage{algorithm}
\usepackage{algorithmic}
\usepackage{booktabs}
\usepackage{threeparttable}

\newtheorem{hypothesis}{Hypothesis}

\usepackage{newfloat}
\usepackage{listings}
\DeclareCaptionStyle{ruled}{labelfont=normalfont,labelsep=colon,strut=off} 
\floatstyle{ruled}
\newfloat{listing}{tb}{lst}{}
\floatname{listing}{Listing}
\title{Measuring Information Richness in Product Images: Implications for Online Sales}
\author {
    Yuting Zhu\textsuperscript{\rm 1},
    Xinyu Cao\textsuperscript{\rm 2},
    Yuzhuo Su \textsuperscript{\rm 2},
    Yongbin Ma\textsuperscript{\rm 3}    
}
\affiliations {
    \textsuperscript{\rm 1} National University of Singapore\\
    \textsuperscript{\rm 2} The Chinese University of Hong Kong\\
    \textsuperscript{\rm 3} Ningbo University 
}

\begin{document}

\maketitle
\begin{abstract}
A common challenge for e-commerce sellers is to decide what product images to display on online shopping sites. In this paper, we propose and validate a novel metric, \emph{k-value}, to quantify the information richness of an image set, and we further investigate its effect on consumers' purchase decisions. We leverage patch-level embeddings from Vision Transformers (ViT) and apply k-means clustering to identify distinct visual features, defining k-value as the number of clusters. An online experiment demonstrates that k-value aligns with human-perceived information richness, validating the metric. A simulated online shopping experiment further reveals a significant yet counterintuitive result: while an image set with a higher k-value (richer information) shortens decision time, it paradoxically reduces purchase propensity. Our findings illuminate the complex relationship between visual information richness and consumer behavior, providing sellers a quantifiable tool for image selection.

\end{abstract}

\section{Introduction}
Product images play a pivotal role in online shopping, as it is the major source from which consumers get product information before making their purchase decision (e.g., \citealt{chenimage}). For example, for fashion products, consumers mainly rely on product images to learn about product features such as their pattern, print, cut, occasion, color, etc. According to a McKinsey report (\citealt{mckinsey}), product images are ranked as one of the most important factors that affect online shopping experience. 

Given the critical role of product images in online shopping, it is important to examine how the amount of information conveyed via product images influences consumers' purchase behavior. Despite its importance, relatively few studies provide direct empirical evidence, primarily due to the challenge of measuring the amount of information contained in a set of images. To address this gap, our paper introduces a novel methodology that quantifies the amount of information contained in images by integrating cutting-edge computer vision techniques with established theories from economics and marketing. 

Specifically, we define the amount of product information conveyed by a set of images as the number of distinct features they present, a perspective grounded in economics and marketing theory (e.g., \citealt{monicmiguel12,ke16}). To operationalize this, we harness the Vision Transformer (ViT; \citealt{vit}), which segments each image into patches and transforms them into vector embeddings. We then pool the patch embeddings across all images in a set and apply an unsupervised clustering algorithm (e.g., K‑means) to group semantically similar patches. The optimal number of clusters, which is determined via standard silhouette scores, serves as our proposed metric of information richness contained in the set of product images. This method integrates state-of-the-art computer vision tools with theoretical constructs of feature-based information in consumer research literature. 

The intuition behind our method is straightforward. Consumers perceive product features through each small segment (i.e., patch) of an image, with context information embedded within those patches. These meaningful visual cues are captured by the ViT’s patch embeddings, which map each patch to a vector representation, that is, a point in a high-dimensional space. By design, similar embeddings reflect similar visual features, and they will be points close to each other in the high-dimensional space. Consequently, grouping patch embeddings via clustering effectively reveals the number of distinct visual features conveyed by the image set. The optimal number of clusters reflects the best way to group these points, and thus serves as our proposed metric of information richness contained by a set of product images. 

We conducted two online experiments. The first experiment serves as an empirical validation of our proposed metric, confirming that our proposed metric aligns closely with human perceptions of visual information richness. Building on this, we conduct a second experiment designed to uncover the relationship between the amount of information contained in product images and consumers' purchase behavior: we randomly generate different image sets with different amount of product information contained, and examine how they affect participants' purchase behaviors. We find that, conditional on the same number of images, a higher level of information richness contained in product images shortens consumers' decision time, but lowers their purchase intention at the same time. A possible explanation is that more information makes consumers more certain about their preferences, but a lower uncertainty in preferences reduces consumers' purchase likelihood on average \citealt{cao2021preference}. Another possible mechanism is that a higher level of information richness may trigger consumers' cognitive overload, thereby reducing their decision time and purchase intention. This is consistent with the classic theory of information overload where excess information overwhelms cognitive capacity and reduces decision satisfaction (\citealt{informationoverload, Eppler01112004}). 

Our contributions are twofold. First, we develop a novel, theoretically grounded method to quantify the amount of information conveyed through product images. We empirically validate our proposed metric against information richness perceived by human participants, confirming its validity as a perceptual proxy. Second, we provide empirical evidence on how the amount of information conveyed in product images affects consumers' purchase behaviors. Counterintuitively, we find that a higher level of image-based product information decreases consumers' purchase likelihood. 

Our code and processed data will be made publicly available upon publication to ensure full reproducibility of our results.

\section{Related Work}
Several measures have been proposed in the literature to quantify the amount of information in images. The first approach is based on Shannon entropy (\citealt{6773024}), a foundational concept in information theory that quantifies the uncertainty or unpredictability of a source. Applied to images, Shannon entropy measures the information content based on the probability distribution of pixel intensities, with higher entropy indicating greater complexity or variability in the image content (\citealt{digitalimageprocessing}). A second approach uses mutual information to capture the amount of shared information between two random variables or signals (\citealt{1057418}). In image analysis, mutual information is commonly employed to assess how much knowledge of one image reduces uncertainty about another. A third approach involves Kolmogorov complexity (\citealt{e13030595}), also known as algorithmic complexity. This measure considers the information content of an individual image in a distribution-free manner and is defined as the length of the shortest computer program that can reproduce the image. Images with highly regular patterns that can be described by compact algorithms exhibit low Kolmogorov complexity.

Our contribution differs in that we focus on quantifying \textit{product information} embedded in images, information that is particularly relevant in consumer decision-making. To this end, we begin with foundational theories from economics and marketing, and operationalize these concepts using computer vision techniques to capture the richness of product-relevant visual information.

A growing body of research has established that online product images play a critical role in shaping consumer behavior. Prior studies have demonstrated that images can build consumer trust (\citealt{magoogleresearch}), influence brand perception (\citealt{liudaria}), affect product demand (\citealt{zhangimage22}), and even impact post-purchase outcomes such as return rates (\citealt{maratreturn}) and restaurant survival rates (\citealt{zhanglan22}). While these studies underscore the importance of images in online marketplaces, they primarily focus on the presence or absence of images or on high-level visual features such as quality, aesthetics, or style. Notably, none of the existing literature examines the role of product information richness embedded in images, namely, the extent to which images convey concrete, attribute-level information about the product itself. Our work fills this important gap by proposing a novel, theoretically grounded and computationally scalable approach to quantifying product information richness in images. By doing so, we move beyond treating images as a mere visual cue and instead conceptualize them as an active medium for product communication, offering new insights into how the content of images influences consumer decision-making.

\section{Methodology}\label{sec:methodology}
\begin{figure*}[h!]
    \centering
    \includegraphics[width=1\linewidth]{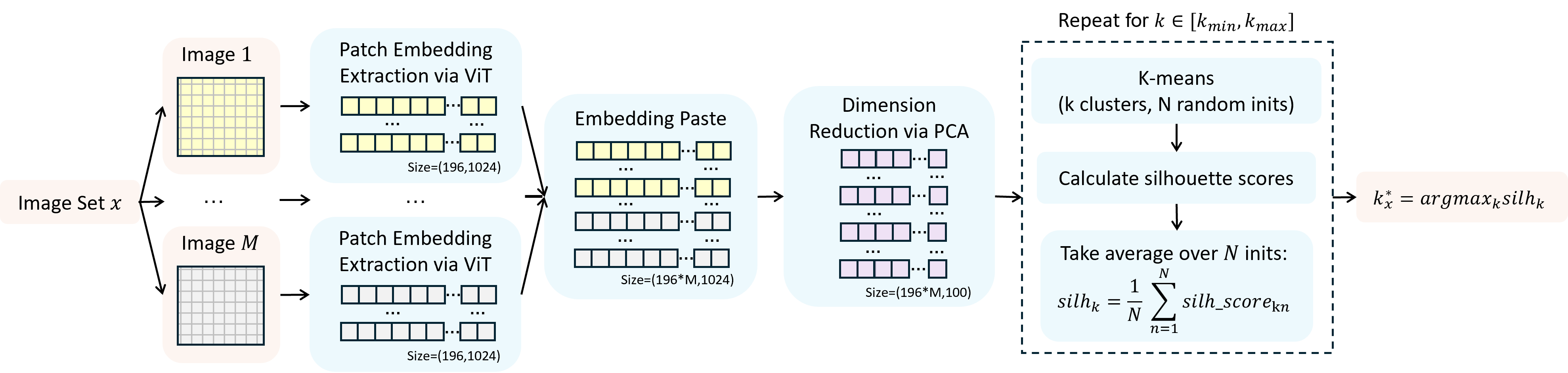}
    \caption{Illustration of Image Information Quantification Process}
    \label{fig:kExtraction}
\end{figure*}

\begin{figure}[h!]
    \centering
    \includegraphics[width=0.75\linewidth]{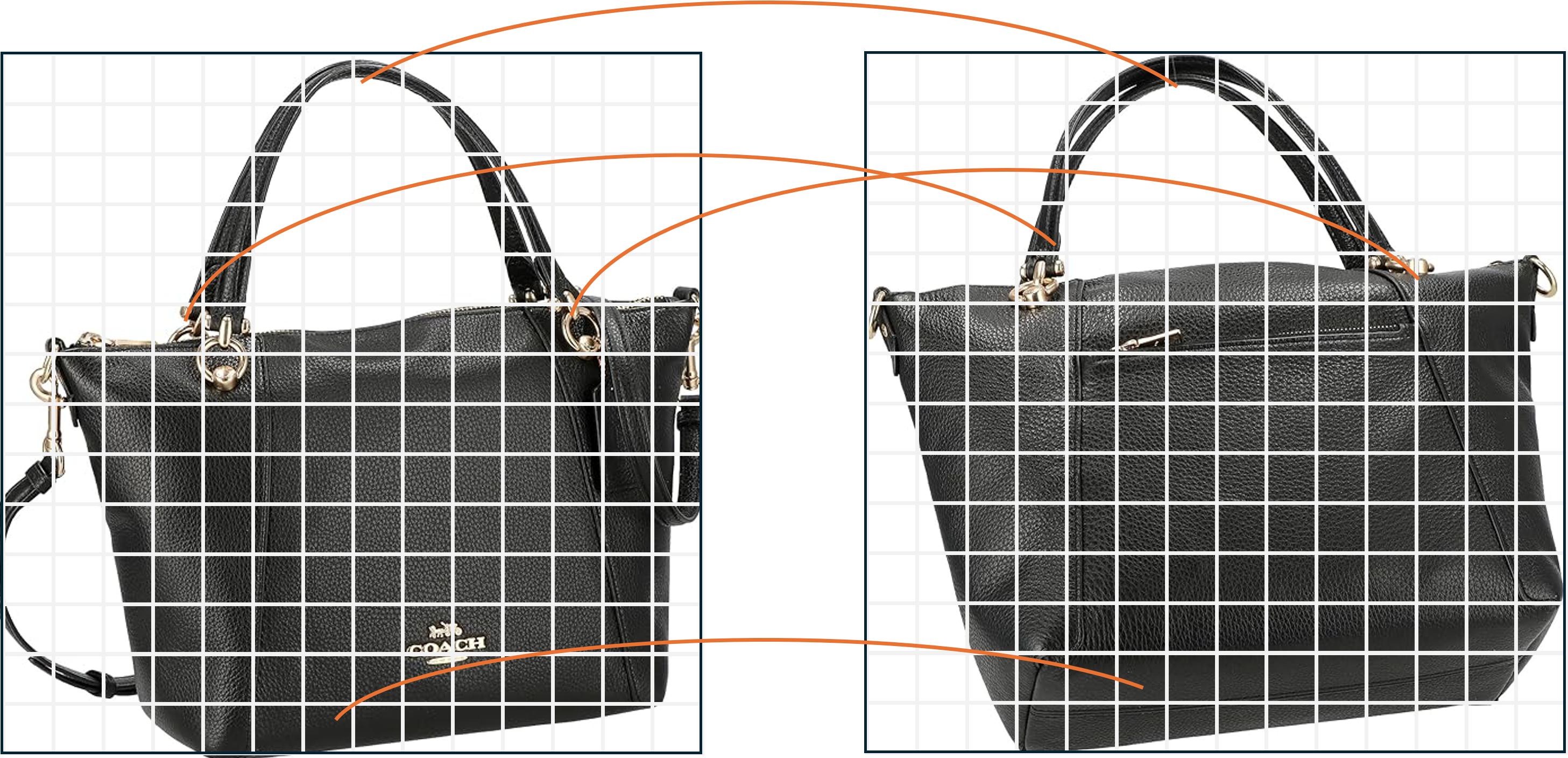}
    \caption{Illustrative example of our approach. The set of images displays the front and back of a handbag. Visual elements that appear in both images, such as the shape, color, handles, and hardware of the bag, are connected by red lines. The ViT's patch embeddings capture these visual cues, and by clustering the embeddings, we can identify the number of distinct visual cues (which can be considered as product features) conveyed by the image set.}
    \label{fig:k_example}
\end{figure}

The overall process to quantify the amount of information in images is summarized in Figure~\ref{fig:kExtraction}, with an illustrative example provided in Figure~\ref{fig:k_example}.

Suppose an image set consists of $M$ images. First, for each image, we extract patch embeddings using a Vision Transformer (ViT) model pretrained on the ImageNet-21k dataset (ViT-L\_16, available at \url{https://github.com/google-research/vision_transformer}). Specifically, we use the embeddings output by the transformer encoder---prior to the MLP head---as our patch representations. The embeddings capture visual cues from each image patch while incorporating contextual information. The resulting output consists of 196 patches per image, with each patch represented by a 1024-dimensional vector. In other words, each image is represented by a $(196, 1024)$ matrix. 

Next, we stack the patch embeddings of all $M$ images in the image set, resulting in a $(196*M, 1024)$ matrix. With each 1024-dimensional vector representing a point in the 1024-dimensional space, this matrix represents $196
*M$ points in a 1024-dimensional space, and our next step is to group them to $k$ clusters according to their location in the space. Note that clustering in a high-dimensional space is computationally challenging. To reduce dimensionality, we apply Principal Component Analysis (PCA) and keep the first $L=100$ principal components, based on preliminary tests that the top 100 components capture 95\% of the data variance on average. The resulting output is a $(196*M, 100)$ matrix, representing $196*M$ points in a 100-dimensional space.

In the final step, we apply the K-means clustering method to the $196*M$ points in the 100-dimensional space. To determine the optimal number of clusters, we use the silhouette score, which evaluates both the cohesion within clusters as well as the separation between clusters, to capture clustering quality. In particular, for each $k$ in a predefined range $[k_{min},k_{max}]$, we perform K-means clustering $N=30$ times with different random initializations. In each run $n\in\{1,2,...,N\}$, we compute the silhouette score ($silh\_score_{kn}$). The average silhouette score over the $N$ runs captures the average clustering quality with $k$ clusters.
\begin{equation}
    silh_k = \frac{1}{N}\sum_{n=1}^Nsilh\_score_{kn}.
\end{equation}
To search for the optimal number of clusters $k$, we increment $k$ by a step size of 3 and set the following stopping rule: the search is terminated if the average silhouette score does not improve over the best recorded value for 100 consecutive iterations. The value of $k$ that achieves the highest average silhouette score among all the searched $k$ is selected as the optimal number of clusters $k^*$.
\begin{equation}
    k^* = argmax_k silh_k.
\end{equation}
While the points grouped in one cluster can be considered as containing similar information, we consider the optimal number of clusters $k^*$ representing the amount of information contained in the image set. Note that while we use K-means clustering method in this paper, our method is compatible with other clustering methods as well. 

\section{Experiments}
\subsection{Experiment 1: Metric Validation}
Experiment 1 evaluates the validity of our proposed metric by examining its alignment with consumers' perception of information richness in image sets. In particular, we test the following hypothesis.
\begin{hypothesis}
For the same product, an image set with a higher value of $k^*$ will be perceived by consumers as providing more product information. 
\end{hypothesis}

\begin{figure}
\includegraphics[width=0.4\linewidth]{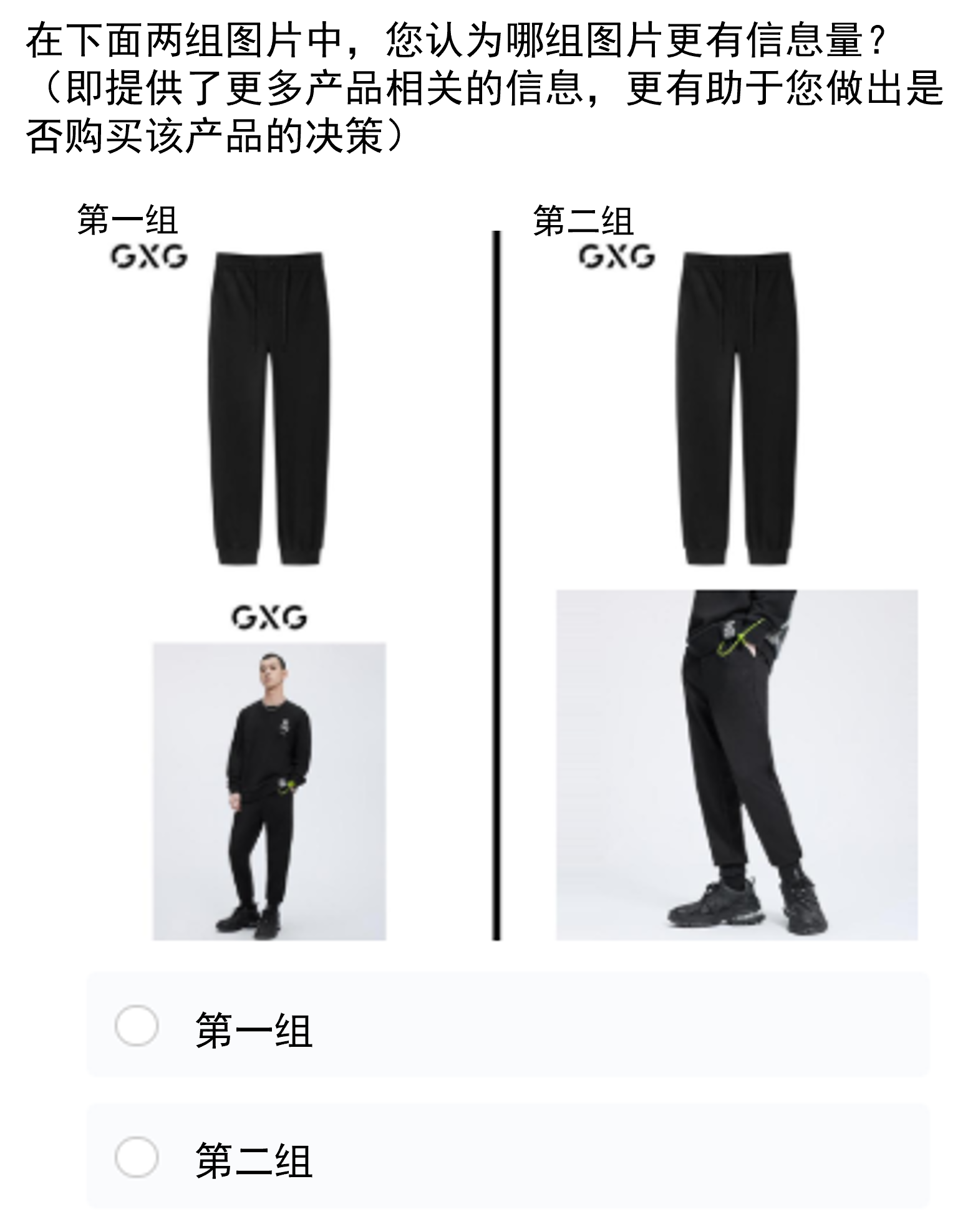} 
\hspace{0.3cm}
\includegraphics[width=0.45\linewidth]{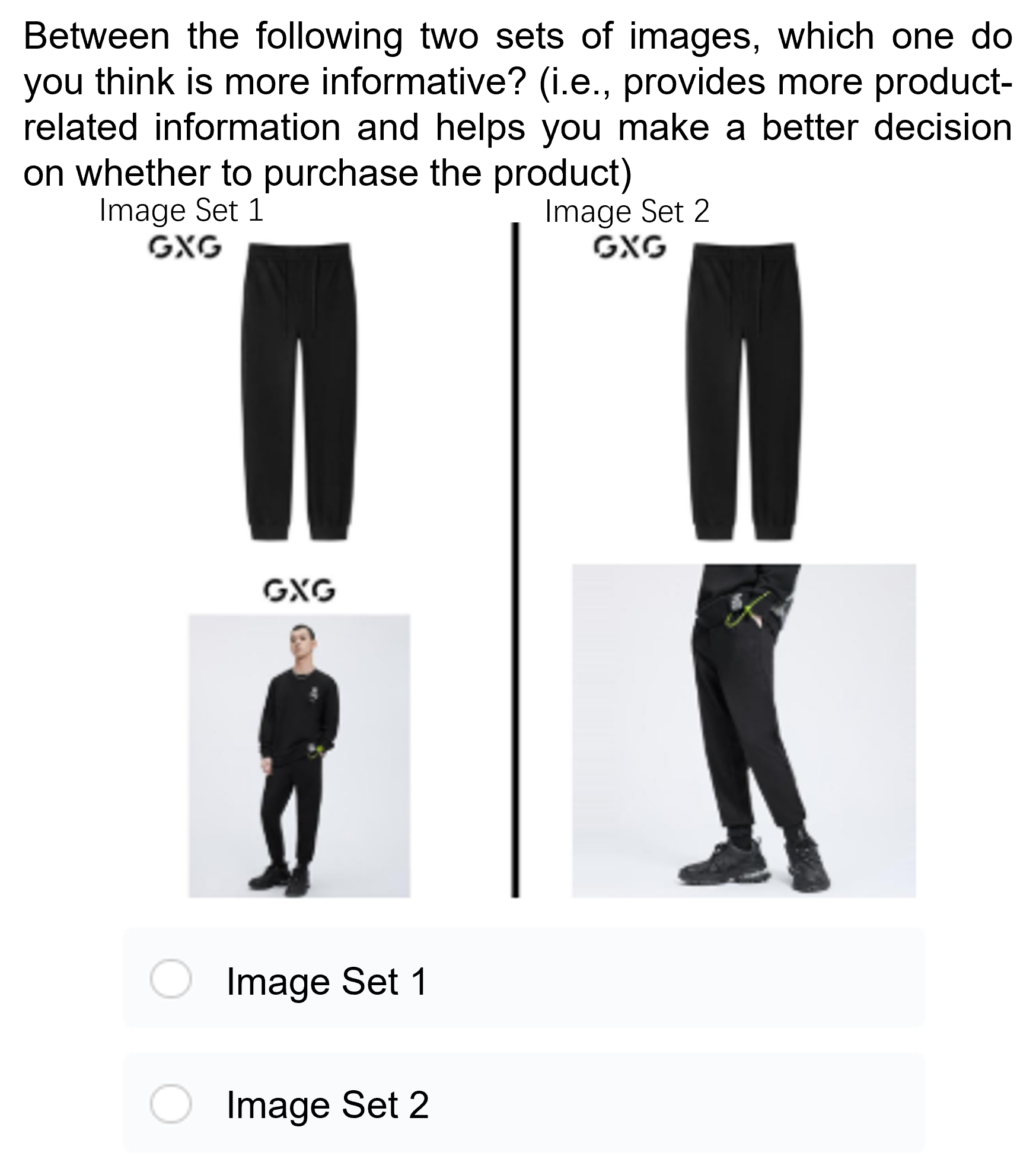} 
\caption{Experiment 1: An Example Question. The left-hand-side shows the original format used in the experiment (in Chinese), and the right-hand-side shows the English translation.}
    \label{fig:Exp1Demo}
\end{figure}

In the online experiment, we constructed 1,000 questions using product images from 1000 distinct products. Each question was randomly assigned to five participants. We recruited 500 participants in total, with each participant answering 10 questions. Each question displayed two image sets depicting the same product, each containing 1-5 images, and participants were asked to select the image set they perceived as providing more product information based solely on visual content. To prevent decisions based purely on image quantity, we constrained the two image sets in each question to contain the same number (but different combinations) of images. The two image sets were displayed side-by-side to facilitate direct comparison. Figure~\ref{fig:Exp1Demo} illustrates an example question.

These 1,000 products were sourced from six distinct fashion brands. All product images used in the online experiment were obtained from a major Chinese e-commerce platform to ensure the stimuli reflected real-world consumer experiences. Data collection was conducted via Credamo\footnote{www.credamo.com}, a reputable Chinese survey platform recognized for high-quality data. Participants received a compensation of 1 Chinese Yuan each. To ensure data quality, we embedded attention-check questions (e.g., simple logic tasks) in the survey. Participants who failed these checks or exited the survey prematurely were excluded. Recruitment continued until 500 participants successfully met our data requirements. In addition, to prevent rushed responses, we enforced participants to spend at least five seconds in each image-set-choice question before moving to the next question.


To test Hypothesis 1, we run the following logistic regression:
\small
\begin{equation} \label{eq:Exp1_logit}
logit(\Pr(Y_{ij}=1))=\beta_0 + \beta_1 \left(k_{j1}^* - k_{j2}^*\right) + \beta_3 \left(X_{j1}-X_{j2}\right) + \epsilon_{ij},
\end{equation}
\normalsize
where $logit(p)=\ln(p/(1-p))$. $Y_{ij} = 1$ means that for product $j$, participant $i$ perceived image set 1 to be more informative than image set 2, and $Y_{ij}=0$ means participant $i$ perceived image set 2 to be more informative. $k^*_{js}$ denotes our proposed metric for image set $s$ ($s=1,2$). $X_{js}$ ($s=1,2$) represent image characteristics, including brightness, contrast, blur, saturation, and colorfulness, clarity score, aesthetic score, degree of black-and-white, and image purity—features extracted using the Tencent Cloud Image Quality API. These measures are all averaged across images in each set. 

Our parameter of interest is $\beta_1$. Suppose Hypothesis 1 is true, then when $k_{j1}^* - k_{j2}^*$ is larger, each participant is more likely to perceive image set 1 as more informative (i.e., $\Pr(Y_{ij}=1)$ increases). Thus, $\beta_1$ should be positive if Hypothesis 1 is true. 

The regression results are shown in Columns (1) and (2) of Table~\ref{tab:Exp1}. Column (1) reports the estimate without controlling for image characteristics, while column (2) includes these controls. 
As a robustness check, we use an alternative independent variable, $k^*_{j1}/(k^*_{j1}+k^*_{j2})$ to replace $k_{j1}^* - k_{j2}^*$ in Equation (\ref{eq:Exp1_logit}). The results are reported in Columns (3) and (4). In all four columns, the estimate of $\beta_1$ is significantly positive. These results support Hypothesis 1, offering robust empirical evidence that our proposed metric provides a valid measure of information richness of an image set. 


\begin{table}[h!]
\centering
\setlength{\tabcolsep}{1mm}
\begin{threeparttable}
\begin{tabular}{lcccc}
\toprule
& \multicolumn{4}{c}{$logit(\Pr(Y_{ij}=1))$} \\
&(1)&(2)&(3)&(4)\\
\midrule
$k^*_{j1}-k^*_{j2}$&0.0013$^{**}$&0.0011$^{**}$&&\\
&(0.0005)&(0.0005)&&\\
$k^*_{j1}/(k^*_{j1}+k^*_{j2})$&&&1.0357$^{***}$&0.9720$^{***}$\\
&&&(0.1574)&(0.1604)\\
$X_{j1}-X_{j2}$ &  & Yes &  & Yes \\
\midrule
Num.Obs.&5000&5000&5000&5000\\
pseudo-$R^2$ & 0.001 & 0.005 & 0.006 & 0.010\\
\bottomrule
\end{tabular}
\begin{tablenotes}
    \footnotesize
\item Notes. Intercept omitted. The standard errors are reported in parentheses. $^{*}p<0.1$, $^{**}p<0.05$, $^{***}p<0.01$.
\end{tablenotes}
\end{threeparttable}
\caption{Experiment 1: Metric Validation}
\label{tab:Exp1}
\end{table}

\subsection{Experiment 2: Implication for Online Sales}
In Experiment 2, we simulate an online shopping site to examine how the amount of information conveyed in product images, quantified by our proposed metric $k^*$, affects consumers' purchase behavior. 

First, we randomly select 100 products priced below 1,000 Chinese Yuan from the product pool of the Chinese e-commerce platform as mentioned in Experiment 1, resulting in 51 male-oriented products and 49 female-oriented products. For each selected product, from all the available product images, we randomly generate two image sets, each with 1-5 images. 

We recruit 1,000 participants through Credamo for the experiment. To match the gender distribution of the product set, 510 participants were male and 490 were female.\footnote{Participants were excluded from the final sample if they failed the attention check questions, did not meet the minimum response time requirement (2 seconds per product), or exited the survey before completion. The resulting dataset comprised 996 valid participants.} For each participant, we randomly select 10 different products, letting them evaluate one product per page. The participant decide whether to purchase or not to purchase each product based on the price, a brief description, and a randomly selected image set (out of the two image sets) of the product. Note that male (female) participants were shown male-oriented (female-oriented) products only. 

\begin{figure}[h!]
\includegraphics[width=0.4\linewidth]{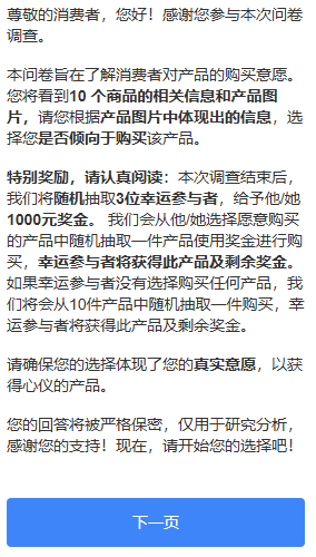} 
\hspace{0.3cm}
\includegraphics[width=0.55\linewidth]{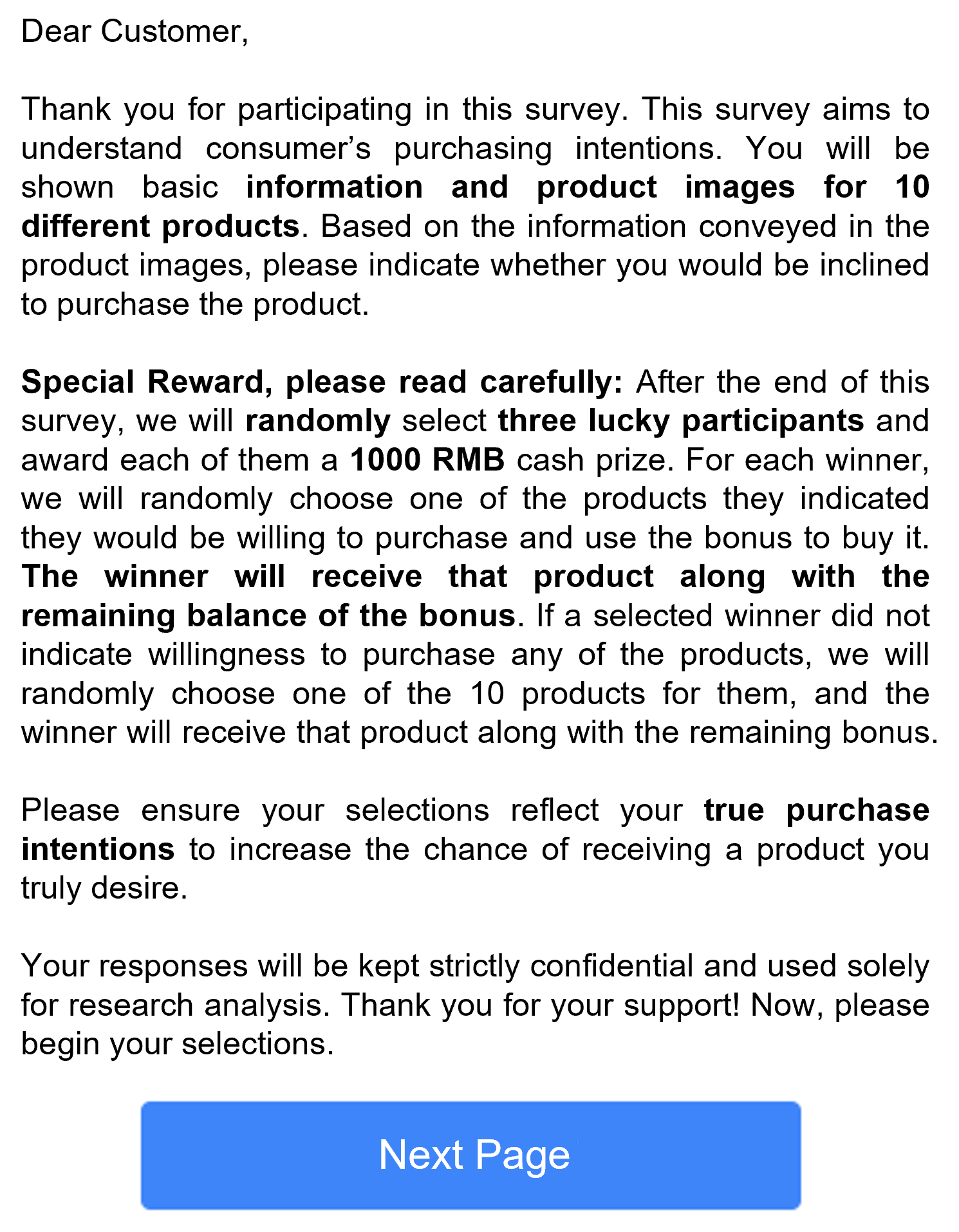} 
\caption{Experiment 2: Experiment Instruction Page. The left-hand-side shows the original format used in the experiment (in Chinese), and the right-hand-side shows the English translation.}
    \label{fig:instruction}
\end{figure}

\begin{figure}[h!]
\includegraphics[width=0.4\linewidth]{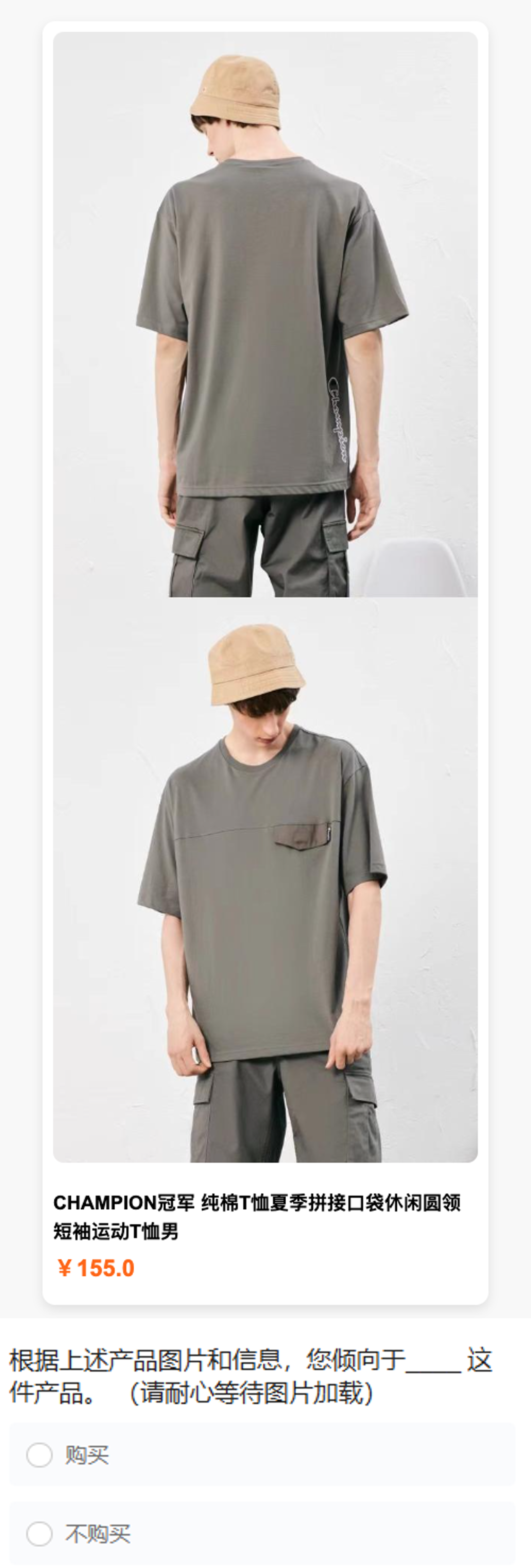} 
\hspace{0.3cm}
\includegraphics[width=0.43\linewidth]{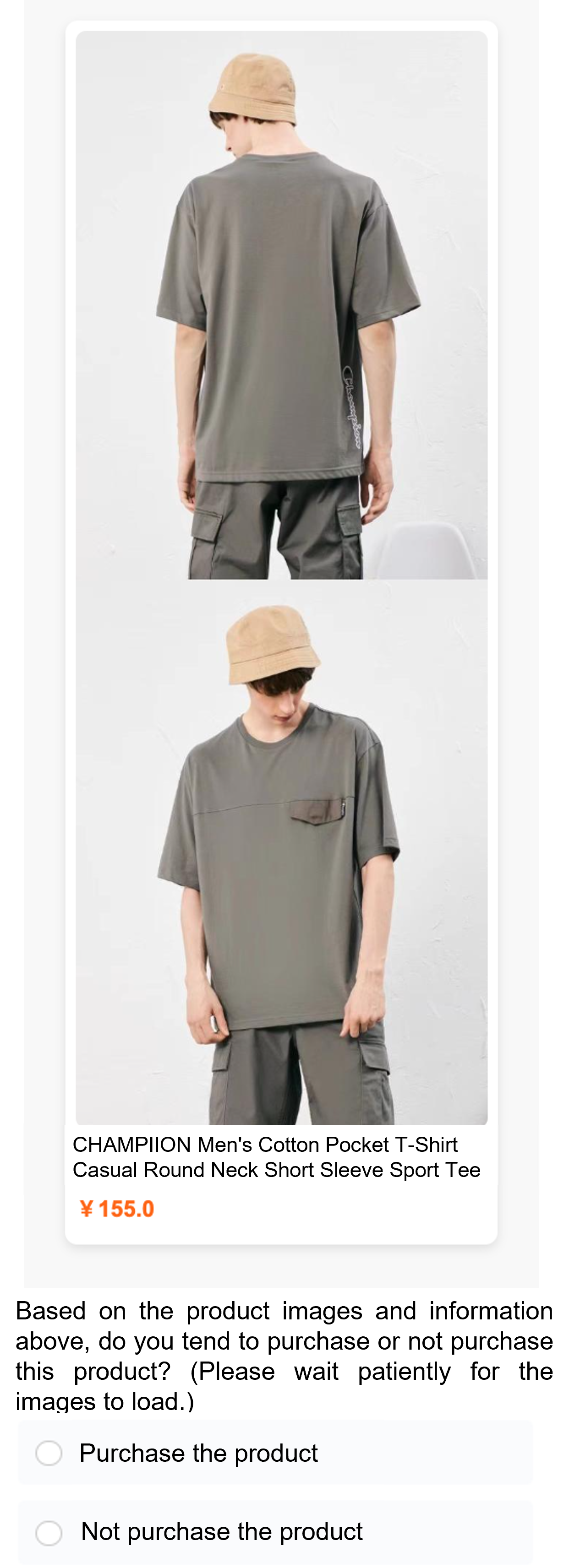} 
    \caption{Experiment 2: An Example Question. The left-hand-side shows the original format used in the experiment (in Chinese), and the right-hand-side shows the English translation.}
    \label{fig:Exp2Demo}
\end{figure}

To encourage participants make the choices carefully and reveal their true purchase intention, before they start answering the questions, we notify them that we will randomly select three lucky participants and award each of them 1000 Chinese Yuan. For each lucky participant, we will use the award to purchase a randomly selected product out of the products they indicated willingness to purchase, and this participant will get the product together with the remaining cash. If the participant did not choose to purchase any of the 10 products, we will randomly choose one of the 10 products to purchase, and the participant will receive the product together with the remaining cash. The experiment instruction page is provided in Figure~\ref{fig:instruction}, and an example of the product page shown to participants is provided in Figure~\ref{fig:Exp2Demo}. 

To examine how the information richness conveyed in product images affect consumers' purchase intention, we estimate the following fixed-effects model:
\begin{equation} \label{eq:purchase}
Purchase_{ijs} =  \beta_1 k_{js}^* + \beta_2 p_j + \beta_3 NImage_{js} +  \alpha_b + \gamma_i + \epsilon_{ijs}.
\end{equation}
The outcome variable $Purchase_{ijs}=1$ if participant $i$ chose to purchase product $j$ when shown image set $s = \{1,2\}$, and $Purchase_{ijs}=0$ otherwise. The key independent variable, $k_{js}^*$, is our proposed metric capturing the information richness of image set $s$ $(s=1,2)$ for product $j$. We control for product price $p_j$ and the number of images  $NImages_{js}$. We further control for brand fixed effects ($\alpha_b$) and participant fixed effects ($\gamma_i$) to account for brand-level and individual-level heterogeneity. $\epsilon_{ijs}$ is the error term. 

Table~\ref{tab:Exp2} Panel A presents the regression results.  examining the effect of visual product information on consumer purchase intention. Across all model specifications, the coefficient of $k_{js}^*$ is significantly negative, meaning that a higher level of information richness contained in product images lowers lowers consumers' purchase intention. 

Furthermore, we use decision time (i.e., how long it took participant $i$ to decide whether to purchase product $j$ or not when shown image set $s$) as the outcome variable to run the same set of regressions, as presented in Table~\ref{tab:Exp2} Panel B. Across all model specifications, the coefficient of $k_{js}^*$ is significantly negative, suggesting that a higher level of information richness reduces the time needed to make purchase decisions. 

Combining the results above, we find that displaying richer product information renders consumers make decisions faster but reduce their purchase intention at the same time. A possible explanation is that revealing more information makes consumers more certain about their preferences, and thus it will take them a shorter period of time to make decisions. At the same time, consumers' purchase likelihood decreases on average when they know their preferences better. 


\begin{table}[h!]
\centering 
\begin{threeparttable}
\begin{tabular}{lccc}
\toprule
Panel A: & \multicolumn{3}{c}{Purchase} \\
&(1)&(2)&(3)\\
\midrule
$k_{ja}^*$&-0.1996$^{**}$&-0.1796$^{**}$&-0.1592$^{*}$\\
&(0.0802)&(0.0816)&(0.0822)\\
$p_{j}$&-0.0862$^{***}$&-0.0738$^{***}$&-0.0856$^{***}$\\
&(0.0198)&(0.0279)&(0.0278)\\
$NImage_{ja}$&0.0286$^{***}$&0.0294$^{***}$&0.0295$^{***}$\\
&(0.0041)&(0.0043)&(0.0043)\\
Brand FE  & & Yes & Yes \\
User FE   & &     & Yes \\
\midrule
Num. Obs. &9960&9960&9960\\
$R^2$&0.008&0.008&0.189\\
\midrule
\midrule
Panel B: & \multicolumn{3}{c}{Decision Time (seconds)} \\
&(1)&(2)&(3)\\
\midrule
$k_{ja}^*$ & -0.1924$^{*}$ & -0.2639$^{**}$ & -0.2579$^{**}$ \\
& (0.1040) & (0.1057) & (0.1093) \\
$p_{j}$  & -0.0169 & -0.0439 & -0.0271 \\
& (0.0257) & (0.0361) & (0.0369) \\
$NImage_{ja}$  & 10.0110$^{*}$ & 5.9263 & 6.2225 \\
& (5.3571) & (5.5666) & (5.7023) \\
Brand FE  & & Yes & Yes \\
User FE   & &     & Yes \\
\midrule
Num. Obs. & 9960 & 9960 & 9960 \\
$R^2$ & 0.001 & 0.003 & 0.142 \\
\bottomrule
\end{tabular}
\begin{tablenotes}
\footnotesize
\item Notes. Standard errors are reported in parentheses. $^{*} p<0.1$, $^{**} p<0.05$, $^{***} p<0.01.$
For convenience of presentation, we have divided $k_{js}^*$ and $p_j$ by 1,000 when including them in the regressions.
\end{tablenotes}
\end{threeparttable}
\caption{Experiment 2: Visual Information Richness and Purchase Behavior}\label{tab:Exp2}
\end{table}

\section{Conclusion}
In summary, our paper offers both methodological and empirical contributions to our understanding of visual product information in online shopping. We introduce a novel, theory-based approach to quantify the amount of information conveyed in product images, leveraging Vision Transformers and unsupervised clustering techniques to capture the number of distinct visual features contained in product images. Our measure is validated against human perceptions, establishing it as a valid proxy of information richness. Using this metric, we conduct a controlled experiment to examine the effect of image-based product information on consumers' purchase behaviors. The results consistently show that, when holding the number of images constant, a higher level of information richness leads to shorter decision time but lower purchase likelihood. 

Future research could extend our framework by exploring interactions between visual and non-visual content. Moreover, further investigation is needed to identify the optimal level of visual product information that balances informativeness with cognitive ease, which can offer actionable insights for platform design and visual merchandising strategies.

\bibliography{imageinformation}

\end{document}